\documentclass[11pt]{article}
\usepackage{graphicx}

\begin{document}

%\baselineskip=17pt

%\parskip=5pt

%\preprint{hep-ph/0512327}
%\preprint{WSU-HEP-0505}
%\hspace*{\fill} $\hphantom{-}$

\begin{center}
{\bf\large{On Two-Body Decays of A Scalar Glueball }}
\vskip 10mm
K.T. Chao$^1$, Xiao-Gang He$^{2,1}$, and J.P. Ma$^{3,1}$
\\
{\small\it{$^1$Department of Physics, Peking University,
Beijing\\
$^2$Department of Physics and Center for Theoretical Sciences,
National Taiwan University,
Taipei\\
$^3$Institute Of Theoretical Physics, Academia Sinica, Beijing}}
\end{center}
%\date{\today}
\vskip 1cm

\begin{abstract}
We study two body decays of a scalar glueball. We show that in QCD a
spin-0 pure glueball (a state only with gluons) cannot decay into a
pair of light quarks if chiral symmetry holds exactly, i.e.,  the
decay amplitude is chirally suppressed. However, this chiral
suppression does not materialize itself at the hadron level such as
in decays into $\pi^+\pi^-$ and $K^+K^-$. We show this explicitly in
two cases with the glueball to be much lighter and much heavier than
the QCD scale using low-energy theorems and perturbative QCD. For a
heavy glueball, using QCD factorization based on an effective
Lagrangian, we find that the hadronization into $\pi\pi$ and $KK$ leads
to a large difference between ${\rm Br} (\pi^+\pi^-)$ and ${\rm
Br}(K^+K^-)$, even the decay amplitude is not chirally suppressed.
Our results can provide some understanding of the partonic contents
if ${\rm Br}(\pi\pi)$ or ${\rm Br}(K\bar K)$ is measured reliably.
\end{abstract}
\vskip 1cm

%\pacs{PACS numbers: }
\par
%\maketitle
%\noindent {\bf Introduction}
\par
It is believed that all hadrons are built with quarks and gluons,
which are the dynamical degrees of freedom of QCD. So far all
observed hadrons have been shown to contain quarks. In general, it
is also possible to have hadrons which contain gluons only, the so
called pure glueball states. Experimentally, the existence of
glueballs has not been confirmed although there are some
indications. Studies with Lattice QCD indicate that the lowest
lying glueball is a scalar, $0^{++}$ state, having a mass in the
range of $1.5\sim 2.0$ GeV\cite{lattice}. The state $f_0(1710)$ is
a promising candidates for a scalar glueball\cite{PDG}.

\par
In the framework of QCD a scalar glueball $G_s$ is in general a
superposition of many components containing gluons and quarks as
partons $a_i(i=1,\cdots, n)$ which can be schematically represented as
\begin{equation}
  \vert G_s \rangle = \sum_{n=2} \psi_{a_1\cdots a_n}
  \vert a_1,\cdots, a_n \rangle = \psi_{gg} \vert gg \rangle+ \psi_{q\bar q}
  \vert q \bar q\rangle +... ,\label{qg}
\end{equation}
where $\psi_{a_1\cdots a_n}$ is the probability amplitude for the
component $\vert a_1,\cdots, a_n \rangle$. It is clear that a
state should not be identified as a glueball state, if it has a
quark content larger than its gluon content, roughly speaking, if
$\vert \psi_{gg} \vert < \vert \psi_{qq} \vert $. Decay products
of a particle can be used to extract crucial information about
whether a state is a glueball or not. In this letter we will show
that two body decays of a scalar glueball can reveal some
important information, and discuss possible experimental
implications. Part of our results, in particular  the results on
pQCD calculation of the leading contribution for glueball decays
into two light mesons have been discussed in Ref.\cite{CHM}. Here
we provide more details including some higher twist effects and
also discussions for low-energy theorem implications for light
glueball decay into two light mesons.
\par
We will first show in QCD, without any assumption, that a $0^{++}$
glueball $G_s$ cannot decay into a light-quark pair $q\bar q$ if
$G_s$ is a pure glueball with exact chiral symmetry. The decay is
chirally suppressed. Then we study the two-body hadronic decays,
such as $\pi \pi$ and $K K$ and show that the quark level chiral
suppression does not materialize itself at hadron level, even for
a pure glueball decay. We will show this explicitly in two cases
with the glueball to be much lighter and much heavier than the QCD
scale. In the case that the glueball is light, the decay products
will have small momenta. One can use low-energy theorems to show
that even in the chiral limit the glueball still can decay into
$\pi\pi$. If the glueball is heavy, one can show based on QCD
factorization even for a pure glueball it will mainly couple to
two quark pairs $q\bar q q\bar q$ which hadronize to two light
mesons or so at long distances rather than just one quark pair
$q\bar q$ at short distances(see Fig.2 (a)). Hence, there is no
chiral suppression for the $\pi\pi$ mode compared with the $KK$
mode. Taking $f_0(1710)$ as an example, we find that a small decay
ratio for $B(\pi^+\pi^-)/B(K^+ K^-)$ does not necessarily imply
that $f_0(1710)$ is a pure glueball. This is in contrast to the
recent result in \cite{Chan}.
\par
%%%%%%%%%%%%%%%% Inset Fig. 1 here %%%%%%%%%%%%%%%%%%%%%%%%%%%%%%

\begin{figure}[hbt]
\begin{center}
\includegraphics[width=8cm]{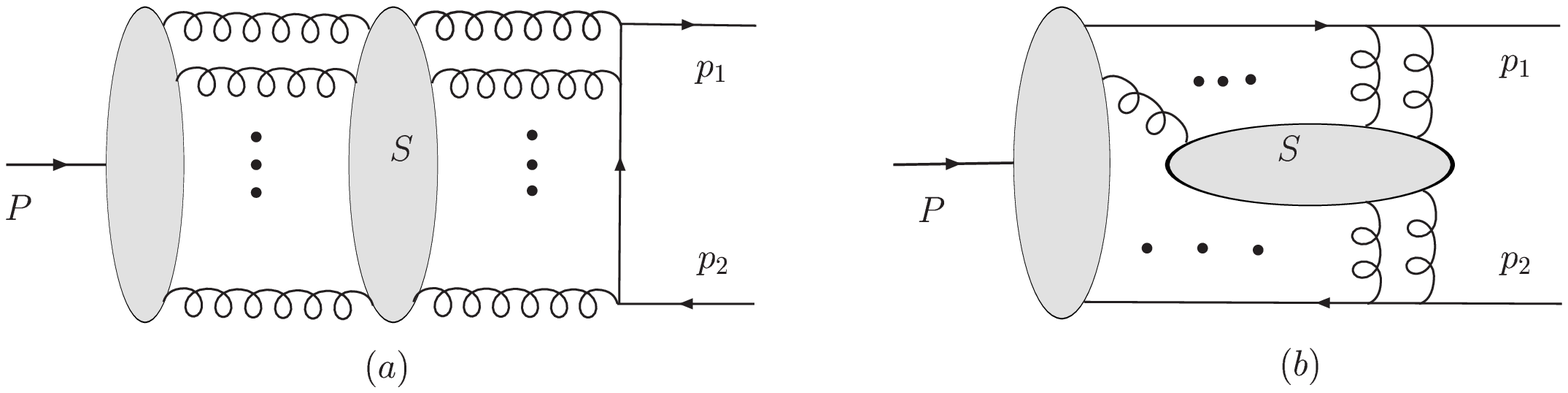}
\end{center}
\caption{A glueball decays into a $q\bar q$ pair. (a) The
contribution from components containing gluons only. (b) The
contribution from components containing a $q\bar q$ and gluons
when mixing exists.} \label{Feynman-dg1}
\end{figure}
\par

%\noindent {\bf The decay of a scalar glueball to a light quark pair}
\par
For the decay of a scalar glueball into $q \bar q$
all components in Eq.(1) may contribute.
The contributions from those components only containing
gluons can be represented by Fig.1a, where the bulb with $S$
can be defined as a $n$-point Green's function of gluon fields
combined with gluon propagators in the free case, and the other bulb attached with
the glueball can
be defined with gluon field operators sandwiched between the vacuum
and the glueball state. Although a complete calculation for the
diagram with the structure given in Fig.1a is not possible at
present, some general conclusions can be drawn by using properties
of QCD and Lorentz covariance.
\par
The decay amplitude from Fig.1a for $G_s \to q(p_1) \bar q(p_2)$
can be written as a product of a spinor pair $\bar u(p_1) $ and
$v(p_2)$ with a product of any number of $\gamma$ matrices
sandwiched between the spinors. Because the quark-gluon coupling
in QCD is vector-like, the number of the $\gamma$-matrices is an
odd number when the quark mass $m_q$ is equal to zero. A product
of $\gamma$-matrices with odd number can be reduced to just one
$\gamma$-matrix. Therefore the amplitude from Fig.1a can always be
written as:
\begin{equation}
{\mathcal T}_g(G_s \to q\bar q) = \bar u(p_1) \gamma_u A^\mu v(p_2).
\end{equation}
Although we cannot obtain an explicit expression for $A^\mu$, we
know from Lorentz covariance that it can be written as
$A^\mu (p_1,p_2) = a_1 p_1^\mu + a_2 p_2^\mu$.
With this it is easy to find that in the chiral limit $m_q =0$, the
contribution to the decay amplitude $G_s \to q \bar q$ from the
pure gluonic components is zero. The result also
applies to a pseudoscalar glueball decays into a $q \bar q$ pair.
\par
It is clear that the contribution of these pure gluonic components
to the decay amplitude in the limit $m_q \to 0$ is
\begin{equation}
{\mathcal T}_g (G_s \to q\bar q)\sim m_q + {\mathcal O} (m_q^3),
\end{equation}
because the helicity of quarks can be flipped with a finite quark
mass $m_q$.
By assuming a specific form of the coupling for a scalar glueball
with two gluons as given in Eq.(4) in the below the result for
$G_s$ is also obtained in Ref.\cite{Chan}, with other assumptions.
We emphasis that the above results can be obtained in QCD without
any assumption. The above result is obtained by an analysis in
perturbative theory. It is well-known that the chiral symmetry not
only can be broken by finite quark masses, but also can be broken
spontaneously, the later is a nonperturbative effect.
Therefore the correct statement about the decay
should be that the decay is not allowed if the chiral symmetry
holds. The $m_q$ in Eq.(3) should not be understood as a current
quark mass, but rather as the scale of chiral symmetry breaking.
The effect of spontaneous breaking of chiral symmetry on the decay
can only be studied with nonoperturbative methods, e.g., in
\cite{Jin} for the case here. Combining the nonperturbative effect
the chiral suppression for the ratio
${\mathcal T}_g (G_s \to u\bar u)/{\mathcal T}_g (G_s \to s\bar s)$
will be not so strong
as suggested by current quark masses ratio $ m_{u}/m_s$.
\par
For contributions from components containing $q\bar q$ pairs with
or without gluons, the situation will be different. The $q\bar q$
in the final state can come from one of the $q\bar q$ pairs
through scattering from the already existing quark contents in the
glueball state as shown in Fig.1b. In this case one cannot
conclude that the contribution from Fig.1b is zero in the limit
$m_q =0$.  The reason is that the glueball can have components
with a $q\bar q$ pair and gluons. If these gluons are in a state
like $J^{PC} =1^{--}$, the $q\bar q$ pair must also be in a
$1^{--}$ state. One can show that the contributions from those
components are not zero in the chiral limit.
\par

%\noindent {\bf The decay of a scalar glueball to two light mesons}
\par
In the above, the results are obtained for the decay of $G_s$ into a
$q\bar q$ pair. For a real decay process one has to work with hadron
states. Unfortunately at present the hadronization mechanism is not
well understood. To study the hadronic decays we will therefore
assume that a scalar glueball dominantly couples to gluons and
quarks via the effective Lagrangian\cite{Chan}:
\begin{equation}
L_s = G_s \left \{ \frac{f_g}{M} G^{a,\mu\nu} G^a_{\ \mu\nu} + f_q
\bar q q \right \} + \cdots. \label{LS}
\end{equation}
where $G_s$ is the extrapolation field of the scalar glueball, $M$
is its mass. $f_g$ and $f_q$ are dimensionless coupling constants.
They are related to those probability amplitude in Eq.(\ref{qg}).
If $G_s$ is a pure glueball, the coupling $f_q$ is chirally
suppressed, i.e., $f_{u,d} << f_s$, or zero if the chiral symmetry
is exact. These couplings are unknown, but important information
about them can be extracted from experiment as we will show later.
\par
\par
If the glueball is light enough, it is easy to show with the above
effective Lagrangian that there is no chiral suppression in the
sense that in the chiral limit the decay of the glueball into
$\pi^+\pi^-$ happens. The decay amplitude through the gluonic
coupling $f_g$ in Eq.(4) is given by
\begin{equation}
{\mathcal T}_g(G_s\to \pi^+ \pi^-) = \frac{f_g}{M} \langle \pi^+ (p_1) \pi^- (p_2) \vert
      G^{a,\mu\nu} G^a_{\ \mu\nu} \vert 0 \rangle .
\end{equation}
This amplitude is nonperturbative. However, there exist some
low-energy theorems which give information about the above
amplitude. In the chiral limit one can show\cite{Shif}:
\begin{equation}
 \langle \pi^+ (p_1) \pi^- (p_2) \vert \left ( -\frac{ \beta_0 \alpha_s}{8\pi}\right )
 G^{a,\mu\nu} G^a_{\ \mu\nu} \vert 0 \rangle = (p_1 +p_2)^2 + {\mathcal O
 }(p^4),
\end{equation}
where $\beta_0 =(11-2n_f/3)$ with $n_f$ the number of light
quarks. This result simply tells that the decay can happen in the
chiral limit. Therefore, there is no chiral suppression if the
glueball is light. Similarly, the direct transmission of $q\bar q$
into $\pi\pi$ can also be fixed\cite{XD}.  The same could also be
obtained by using a chiral realization of $L_s$ as described in
\cite{CH}. One can also work out similar expressions for $G_{s}
\to KK$ amplitude.

To show that whether there is a chiral suppression in $G_s\to
\pi\pi$ compared with $G_s \to KK$, one needs to consider the
direct hadronization of $G_s \to q \bar q$ to $G_s\to \pi\pi (KK)$
and also some other possible contributions\cite{chanowitz2}. The
above discussion indicate that $G_s \to q \bar q$ direct
hadronization will not produce chiral suppressions.

\par
We note that the glueball mass is expected to be around 2GeV.
Practically, the applicability of the low-energy theorems is
questionable at this scale. At this energy scale, perturbative QCD
may make some reliable predictions, such as those of the decay of
$\tau$-lepton. Therefore one can employ QCD factorization for
exclusive processes suggested long time ago in Ref.\cite{BrLe},
where the hadronization is parameterized with light-cone wave
functions. In the following we will consider if there is chiral
suppression from pQCD point of view. We will use QCD factorization
with $L_s$ to study the decay $G_s \to \pi^+\pi^-$ in the
following.
\par
%%%%%%%%%%%%%%%% Inset Fig. 2 here %%%%%%%%%%%%%%%%%%%%%%%%%%%%%%

\begin{figure}[hbt]
\begin{center}
\includegraphics[width=8cm]{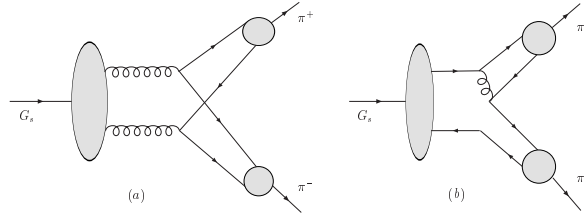}
\end{center}
\caption{(a) One of the 2 diagrams for the decay through the
coupling with gluons.
(b) One of the 4 diagrams for the decay
through the coupling with quarks.} \label{Feynman-dg2}
\end{figure}
\par
We first discuss the contributions from the coupling with gluons.
To the leading twist-2 order, the contribution
comes from diagrams represented by Fig. 2a.
A direct calculation gives:
\begin{eqnarray}
&&{\mathcal T}_g = - \alpha_s  f_g \frac{ 8\pi}{9M} f_\pi^2
\int_0^1 du_1 du_2 \phi_{\pi^+} (u_1) \phi_{\pi^-} (u_2)
\nonumber\\
&&\times \left ( \frac{1}{u_1 u_2} + \frac{1}{(1-u_1)(1-u_2)}
\right )\left [ 1 + {\mathcal O} (\alpha_s) +{\mathcal O}
(\lambda/M)\right],
\end{eqnarray}
where $\phi_\pi$ is the twist-2 light-cone wave function of
$\pi$. $u_i(i=1,2)$ is the momentum fraction carried by
the anti-quark in the meson. In the above, $\lambda$ can be any
soft scale, such as quark mass, $\Lambda_{QCD}$ and $m_\pi$. The
contribution from the coupling with quarks are nonzero if one
takes $m_q\neq 0$.  Clearly, ${\mathcal T}_g$ is not zero in the
chiral limit $m_q=0$.
\par
The contribution from the coupling with quarks is given
by diagrams represented in Fig.2b. It is zero if we only
take twist-2 light-cone wave functions.
At twist-3 there are two wave functions, but only one leads to a
nonzero contributions.
It gives:
\begin{eqnarray}
&&{\mathcal T}_q =-\left . \frac{4\pi}{9} \frac{f_\pi^2}{M^2}
\alpha_s(\mu) \int_0^1 du_1 du_2 \right \{
\phi_{\pi^+}(u_1)\phi_{\pi^-}(u_2)   \left [
   m_u f_u \left (\frac{1}{u_1^2(1-u_2)}
  +\frac{1}{u_1(1-u_2)^2} \right )
\right.
\nonumber\\
 && \left.
            +m_d f_d  \left (\frac{1}{(1-u_1)^2 u_2} +\frac{1}{(1-u_1) u_2^2} \right ) \right ]
 + \frac{m_\pi^2}{m_u + m_d}
 \left [ \left (
      \frac{3-u_2}{u_1(1-u_2)^2} f_u \right. \right.\nonumber\\
      &&\left . + \frac{2+u_2}{(1-u_1) u_2^2} f_d \right )
      \phi_{\pi^-} (u_2) \phi_{\pi^+}^{[p]}(u_1)
\left.\left. + \left (\frac{2+u_1}{u_1^2(1-u_2)} f_u
  + \frac{3-u_1}{(1-u_1)^2 u_2} f_d \right )
      \phi_{\pi^-}^{[p]} (u_2) \phi_{\pi^+}(u_1) \right ] \right
      \}.
\end{eqnarray}
$\phi_{\pi}^{[p]}$ is the twist-3 light-cone wave function.
Definitions of above light-cone wave functions can be found in
\cite{BrFi}.
It should be noted that the above integration is divergent because
of end-point singularities. This is common in an higher-twist
calculation for exclusive processes, examples can be found in
$B$-decay and form-factors\cite{End}. These singularities can be
regularized as usual by introducing a cut-off scale $\Lambda_c$ or
$\epsilon = \Lambda_c/M$ and by changing the integration range
from $[0,1]$ to $[\epsilon, 1-\epsilon]$. In our later discussions
we will use the QCD scale $\Lambda_c = 300$ MeV for illustration.
\par
The amplitude for $G_s \to K^+K^-$ decay can be obtained by
replacing quantities related to  $\pi$ by those related to $K$ correspondingly.
We now apply the above results to
analyze $\pi^+\pi^-$ and $K^+K^-$ decays of $f_0(1710)$ which is a
candidate for a scalar glueball. For numerical calculations we
take the models for twist-2 light-cone wave functions at the
energy scale $1$GeV in \cite{Ball} and the asymptotic form of
$\phi^{[p]}$, which is $1$,
% and  neglect the
% effect from the evolution of the renormalization scale $\mu$ of
%these wave functions.
and take $M=1710$ MeV, $m_u=m_d =4.5{\rm MeV}$, $m_s=120{\rm
MeV}$, $f_\pi = 132$ MeV and $f_k =1.27f_\pi$. We have the
amplitudes in unit of GeV with $\Lambda_c = 300$ MeV:
\begin{eqnarray}
&& {\mathcal T}(\pi^+\pi^-) \approx (-1.062 f_g
-0.602f_u-0.602f_d)\alpha_s \mbox{(GeV)},
\nonumber\\
&& {\mathcal T}(K^+K^-)  \approx (-1.796f_g
-1.674f_u-1.671f_s)\alpha_s \mbox{(GeV)}.\label{twist}
\end{eqnarray}
With smaller cut-off, $T_q$ becomes bigger. The qualitative
features do not change very much.
\par
We note the difference in the coefficients in front of $f_g$ for
the amplitude of $\pi\pi$ and $K\bar K$ in Eq.(\ref{twist}). This
is mainly due to the difference between $f_\pi$ and $f_K$. This
tells that the decays into $\pi\pi$ and $K\bar K$ is already
significantly different, even if the glueball does not couple to
$q\bar q$, i.e., $f_q=0$. With $f_q=0$ the ratio $R={\rm Br}
(\pi^+\pi^-)/{\rm Br}(K^+K^-)\approx {f_\pi^4/f_K^4} = 0.48$,
which is substantially different from 1. This suppression is much
milder compared with the one at the quark level. It should be
noted that the result $R\approx {f_\pi^4/f_K^4}$ can be derived
without the effective Lagrangian in Eq.(4) if the glueball is
purely composed of gluons and the pQCD contribution dominates.
This is because that for a pure gluball state, the amplitude of
the decay $G_s\to \pi^+\pi^-$ can always be written with QCD
factorization as $T_{\pi\pi}= f_\pi^2 H_g \otimes \phi_{\pi^+}
\otimes \phi_{\pi^-}$, where the higher-twist effects related to
$\pi$'s are neglected and $H_g$ consists of some perturbative
coefficient functions and some quantities related to the structure
of $G_s$. $H_g$ does not depend on the hadrons in the final state.
Although $H_g$ is unknown, one can easily find the result of $R
\approx {f_\pi^4/f_K^4}$. Hence, even the decay amplitude is not
chirally suppressed, the difference of hadronization for the
$G_s$-decays into $\pi\pi$ and $KK$ already leads to a large
difference between ${\rm Br} (\pi^+\pi^-)$ and ${\rm Br}(K^+K^-)$.
It should be noted that $R\approx 0.48$ is close to the recent
experimental central value $0.41^{+0.11}_{-0.17}$ obtained by
BES\cite{bes2}. From Eq.(\ref{twist}) the terms proportional to
$f_g$ are sizeable compared with other terms if $f_g$ and $f_q$
are similar in size. Since a glueball should have a larger gluon
content than quark content, $f_q$ should not be too much larger
than $f_g$ if $f_0(1710)$ can be identified as a glueball.

\begin{figure}[htb]
\begin{center}
\scalebox{0.7}{\includegraphics{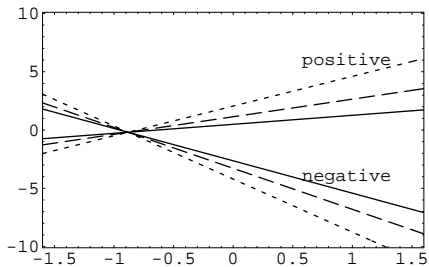}}
%\hspace{0.5cm}\scalebox{0.8}{\includegraphics{ff01.eps}}
\end{center}
\caption{The solid, long-dashed, and short-dashed lines are for
$f_s/f_g$ vs. $f_u/f_g$ with $R= 0.2,\;0.1,\;0.05$, respectively.
Lines labelled with positive and negative are according to the
sign of $\mathcal{T}(\pi^+\pi^-)/\mathcal{T}(K^+K^-)$. }
\end{figure}
\par
If the ratio $R$ is significantly smaller than $f^4_\pi/f^4_K$, it
is an indication that there are other non-gluon content in it.
Previous measurements\cite{PDG} gave smaller values compared with
recent BES data\cite{bes2}. We therefore also studied the
influence of a non-zero $f_q$ on $R$. In Fig.3, we show the
correlation of $f_u/f_g$ and $f_s/f_g$ for several given values of
$R$, where we assume $f_u =f_d$. From Fig.3. we can see that the
measured ratio $R=0.2$ does not necessarily imply $f_u/f_s << 1$,
or the chiral suppression, as discussed after Eq.(4). Experimental
data on $R$ can be explained even if $f_u$ is at the same order of
magnitude as $f_s$, e.g., $f_s/f_g \approx 2f_u/f_g \approx 1$.
Since the couplings $f_q$ are determined by quark contents, the
current experimental data does not exclude the possibility that
$f_0(1710)$ has large quark contents. Combining experimental data
of decays and production in radiative decay of $J/\psi$, the study
in \cite{CZ} also shows that $f_0(1710)$ not only has gluon
content but also large $s\bar s$-content and sizeable $u\bar u +
d\bar d$-content. With the effective Lagrangian $L_s$ one can also
approximate the total decay width to be $\Gamma = \Gamma(gg) +
\sum_q \Gamma(q\bar q)$. If we take the ratio $R$ to be known, the
branching ratio of ${\rm Br}(K\bar K)$ can be expressed as a
function of $f_s/f_g$ or $f_u/f_g$.  In Fig.4 we show the
branching ratio as a function of $f_s/f_g$ for several different
$R$. Reliable experimental data on the branching ratios can
provide crucial information about the constituent contents in
$f_0(1710)$.

\begin{figure}[htb]
\begin{center}
\scalebox{0.7}{\includegraphics{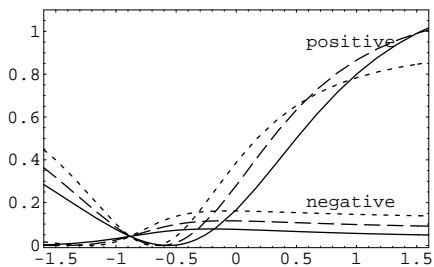}}
%\hspace{0.5cm}\scalebox{0.8}{\includegraphics{brkk03m.eps}}
\end{center}
\caption{The branching ratio of the decay into $K^+K^-$ as a
function of $f_s/f_g$ with cut-off $\Lambda_c = 0.3$ GeV.}
\end{figure}
\par
Our results are different from those in \cite{Chan}. In
\cite{Chan} it is assumed that the decays of $G_s$ into two light
mesons goes like the following, $G_s$ first decays into a $q\bar
q$ pair and then the pair is hadronized into the two light mesons.
Because the decay amplitude into one $q\bar q$ pair is chirally
suppressed, it can result in the chiral suppression at hadron
level. The hadronization is a complicated process, one should not
take directly the quark level picture. We have shown that if the
glueball is light, the low-energy theorems tells us that there is
no chiral suppression. For a heavy glueball, one can use pQCD to
study its decay. In this case the two quark decay picture is also
problematic. In general one needs at least two $q\bar q$ pairs to
form two light mesons. Perturbatively another $q\bar q$ pair can
be produced, e.g., through emission of an extra gluon from the
quark line in Fig.1a and the gluon annihilates into the pair. In
this case the decay amplitude into two $q\bar q$ pairs is not
chirally suppressed.
\par
Using the methods in previous discussions, the coupling of $G_s$
to a proton-antiproton system can also be studied. The coupling is
fixed at certain level by trace-anomaly, the $\sigma$-term and the
strange-quark content of proton. With this approximation we have
considered the possibility if the enhancement in $J/\psi \to
\gamma p \bar p$ at BES\cite{BES} is due to a glueball. We find
that the possible state $X(1876)$ causing the enhancement is
unlikely a scalar glueball\cite{BES}. The coupling of a
pseudoscalar glueball with $p\bar p$ can also be related to the
spin content of the proton as an approximation\cite{BALi}.
Detailed analysis of the coupling to a $p\bar p$ system will be
presented elsewhere.
\par
%\noindent {\bf Conclusions}

In conclusion, we have studied several two body decay modes of a
scalar glueball. Without any assumption we have shown that a pure
spin-0 glueball can not decay into a $q\bar q$ pair in QCD if the
chiral symmetry is exact. Hence the decay is chirally suppressed.
However, this chiral suppression does not materialize itself at
the hadron level such as in $G_s \to \pi^+\pi^-$ and $G_s\to K^+
K^-$. This can be shown in the two cases with the glueball is much
lighter and much heavier than the QCD scale. One expects that the
decay amplitude should not have drastic changes in between and
therefore that the chiral suppression is unlikely materialized in
some intermediate range of the glueball mass. Using QCD
factorization based on an effective Lagrangian for scalar glueball
coupling to two gluons and a quark pair, we have found that even
if the decay amplitude is not chirally suppressed, only from the
difference of hadronization into $\pi\pi$ and $KK$, it already
leads to a large difference between ${\rm Br} (\pi^+\pi^-)$ and
${\rm Br}(K^+K^-)$. The current experimental data of a small ratio
${\rm Br}(\pi^+\pi^-)/{\rm Br}(K^+K^-)$ for $f_0(1710)$ does not
necessarily imply that $f_0(1710)$ is a pure glueball, but it also
allows a sizable $q\bar q$ content. The gluon and quark contents
of $f_0(1710)$ can be better understood if reliable ${\rm
Br}(\pi^+\pi^-, K^+ K^-)$ are measured.

\par\vskip20pt
\noindent {\bf Acknowledgments:}
\par
We thank helpful discussions with Profs. H.-Y. Cheng, X.D. Ji,
H.Y. Jin, Y.P. Kuang and H.-n. Li. This work was supported in part
by grants from NSC and NNSFC (10421003).


\begin{thebibliography}{99}

\bibitem{lattice}
%E. Berg and A. Billorie, Nucl. Phys. B221 (1983) 109; G. Bali, et
%al., (UKQCD Collaboration), Phys. Lett. B309 (1993) 378, C.
%Michael and M. Teper, Nucl. Phys. B314 (1989) 347;
C. Morningstar
and M.J. Peardon, Phys. Rev. D56 (1997) 3043, Phys. Rev. D60
(1999) 034509; C. Liu, Chin. Phys. Lett. {\bf 18} (2001) 187,
Commun. Theor. Phys. {\bf 35} (2001) 288.

\bibitem{PDG} S. Eidelman et al. Particle Data Gropu, Phys. Lett. {\bf B592}, 1
(2004); W.-M. Yao et al. Particle Data Group, J. Phys. G33,
1(2006).

\bibitem{Chan} M.S. Chanowitz, Phys. Rev. Lett. {\bf 95} (2005) 172001, hep-ph/0506125.

\bibitem{CHM} K.-T. Chao, X.-G. He and J.P. Ma, Phys.Rev.Lett. {\bf 98} (2007) 149103, arXiv:0704.1061.

\bibitem{BrLe} S.J. Brodsky and G.P. Lepage, Phys. Rev. D24 (1981) 2848, Phys. ReV. {\bf D22} (1980) 2157.


\bibitem{Jin} Z.F. Zhang and H.Y. Jin, hep-ph/0511252.

\bibitem{Shif} M.A. Shifman,  Phys. Rept. {\bf 209} (1991) 341.

\bibitem{XD} X.D. Ji, Phys. Rev. Lett. {\bf 74} (1995) 1071, Phys. Rev. D52 (1995) 271.

\bibitem{CH} J. Gunion, H. Habar, G. Kane and S. Dawson, The
Higgs Hunter's Guide, Addison-Wesley Publishing Company (1990), J.
Donoghue, E. Golowich and B. Holstein, Dynamics of the Standard
Model, Cambridge University Press (1992), X.-G. He, J. Tandean and
G. Valencia, Phys. Lett. {\bf B631}, 100(2005)[hep-ph/0509041].

\bibitem{chanowitz2} %\cite{Chanowitz:2007ks}
%\bibitem{Chanowitz:2007ks}
  M.~Chanowitz,
  %``Chanowitz replies,''
  Phys.\ Rev.\ Lett.\  {\bf 98}, 149104 (2007).
  %%CITATION = PRLTA,98,149104;%%



\bibitem{BrFi} V.M. Braun and I.B. Filyanov, Z. Phys. C48 (1990) 239.

\bibitem{Ball} P. Ball, JHEP 9901:010,1999, hep-ph/9812375,
P. Ball and M. Boglione, Phys. Rev. {\bf D68} (2003) 094006,
hep-ph/0307337, P. Ball and R. Zwicky,  hep-ph/0510338.

\bibitem{End} M. Beneke et al., Nucl. Phys. B606 (2001) 245,
Z.Z. Song and K.T.  Chao, Phys. Lett. B568 (2003) 127,
hep-ph/0206253, J.P. Ma and Z.G. Si, Phys. Rev. D70 (2004) 074007,
hep-ph/0405111.

\bibitem{bes2} M. Ablikim et al. (BES Collaboration), Phys. Lett.
{\bf B642}, 441(2006).

\bibitem{CZ} F. Close and Q. Zhao, Phys. Rev.  D71 (2005) 094022;
S. Narison, hep-ph/0512256.


\bibitem{BES} J.Z. Bai et al., BES Collaboration, Phys. Rev. Lett. {\bf 91} (2003) 022001.

\bibitem{BALi}B.A. Li, Phys.Rev. D74 (2006) 034019, hep-ph/0510093.

\end{thebibliography}
\end{document}